\documentclass[%
 aps,
 prl,
 amsmath,amssymb,
 showpacs,
 reprint,%
]{revtex4-1}

\usepackage{graphicx}% Include figure files
\usepackage{dcolumn}% Align table columns on decimal point
\usepackage{bm}% bold math
\usepackage{verbatim}
\begin{document}

\title{Ferromagnetism and Borromean Binding in Three-Fermion Clusters}
%\thanks{Footnote to title of article.}

\author{Pavel Kornilovitch}
 \email{pavel.kornilovich@hp.com}
 \affiliation{Hewlett-Packard Company, Printing and Personal Systems, Corvallis, Oregon 97330 USA} 

\date{Received 15 December 2013; revised manuscript received 21 January 2014; published 19 February 2014}  % It is always \today, but any date may be explicitly specified

\begin{abstract}

A three-particle spin-$\frac{1}{2}$ fermion problem with on-site repulsion and nearest-neighbor attraction is solved on the two-dimensional square lattice by discretizing a Schr\"odinger equation in momentum space. Energies of bound complexes (trions) and their binding conditions are obtained. For total spin $S = 1/2$, a wide region of trion instability toward decaying into a stable singlet pair plus a free fermion is identified. The instability is attributed to the formation of a wave function node upon addition of the third fermion. In the $S = 3/2$ sector, trions are found to form in the absence of bound pairs indicating Borromean binding. In the strong coupling limit the system transitions from an $S = 1/2$ ground state to a ferromagnetic $S = 3/2$ ground state in agreement with the Nagaoka theorem for a four-site plaquette.         

\end{abstract}

\pacs{75.10.Jm, 37.10.-x, 71.10.-w}     % PACS, the Physics and Astronomy Classification Scheme

%\keywords{???}               % Use showkeys class option if keyword display desired

\maketitle

{\em Introduction.}---  
The Nagaoka theorem \cite{Nagaoka1966} states that the ground state of a nearly half-filled large-$U$ Hubbard model on the two-dimensional square lattice is ferromagnetic. This property is typically associated with a macroscopic number of fermions. In this Letter, we point out that ferromagnetism can exist in a system of as few as three fermions on an infinite lattice if particle interaction includes a strong enough nearest-neighbor attraction $V$. When both $U$ and $V$ are large compared with intersite hopping $t$, the three particles are confined to an elementary four-site plaquette with one site always left unoccupied. That creates an effective hole and, according to Nagaoka, ferromagnetic alignment of spins.     

Motivated by this expectation, a three-fermion $UV$ model on the square lattice is analyzed here by transforming the Schr\"odinger equation into a set of coupled two-dimensional integral equations~\cite{Rudin1986,Kornilovitch2013}. Both total spins $S = 1/2$ and $S = 3/2$ are studied. An $(S = 1/2) \rightarrow (S = 3/2)$ transition indeed takes place in the $U,V \gg t$ limit. In addition, $S = 3/2$ three-fermion bound states (trions) are found to form at smaller $V$ than triplet pairs, indicating Borromean binding in the $S = 3/2$ sector~\cite{Zhukov1993,Volosniev2013}. In the $S = 1/2$ sector, a wide region of trion instability toward decaying into a singlet bound pair plus a free fermion is identified. The obtained results are relevant to experiments on optical lattices with few atoms~\cite{Jaksch2005,Bloch2008} and to local-pair scenarios of superconductivity~\cite{Micnas1990,Alexandrov1994,Alexandrov2010}.

{\em The model.}--- 
The $UV$ Hamiltonian is given by    
\begin{eqnarray}
H = & - & t \sum_{{\bf m}, {\bf b}, \sigma} 
c^{\dagger}_{{\bf m} \sigma} c_{{\bf m} + {\bf b}, \sigma} + 
\frac{U}{2} \sum_{\bf m} {\hat n}_{\bf m} ( {\hat n}_{\bf m} - 1 ) 
\nonumber \\
    & - & \frac{V}{2} \sum_{{\bf m}, {\bf b}} {\hat n}_{\bf m} {\hat n}_{{\bf m} + {\bf b}}  \: . 
\label{3UV2d:eq:one}
\end{eqnarray}
Here, $c$ and $c^{\dagger}$ are spin-$\frac{1}{2}$ fermion operators, ${\bf m}$ numbers lattice sites, ${\bf b} = (\pm {\bf x}, \pm {\bf y})$ numbers the four nearest site neighbors, $\sigma = \pm \frac{1}{2}$ is the $z$-axis spin projection, and ${\hat n}_{\bf m} = \sum_{\sigma} c^{\dagger}_{{\bf m} \sigma} c_{{\bf m} \sigma}$ is the total particle number operator on site ${\bf m}$. Although Hamiltonian (\ref{3UV2d:eq:one}) is well defined for arbitrary $U$ and $V$, the present Letter is focused on the $U > 0, V > 0$ domain.   

The two-fermion problem (\ref{3UV2d:eq:one}) has been considered by several authors~\cite{Lin1991,Petukhov1992,Kornilovitch2004}, and its main features are summarized below. The discrete singlet spectrum consists of an extended $s$-wave pair and a $d$-wave pair. In general, binding conditions depend on the total pair momentum ${\bf K}$. At ${\bf K} = (0,0)$, the $s$ pair forms at
\begin{equation}
V > V_s = \frac{2 U t}{U + 8t} \: ,
\label{3UV2d:eq:two}
\end{equation}
and the $d$ pair forms at $V > V_d = 2 \pi t/(4 - \pi) = 7.32 \, t$. The energies follow from a $(3 \times 3)$ secular determinant, see Supplemental Material \cite{SupplMat} and Ref.~\cite{Kornilovitch2004}. The discrete triplet spectrum consists of two $p$-wave pairs that both form (at zero momentum) when     
\begin{equation}
V > V_p = \frac{2 \pi t}{\pi - 2} = 5.50 \, t \: . 
\label{3UV2d:eq:three}
\end{equation}

The three-fermion problem (\ref{3UV2d:eq:one}) is much more complicated and has not been considered before. The only relevant study known to the author is due to Rudin~\cite{Rudin1986} who considered a three-boson $UV$ model (\ref{3UV2d:eq:one}), reduced the problem to a set of integral equations using a method identical to ours, but did not attempt numerical solution, perhaps due to lack of adequate computational resources at the time.

\begin{figure*}[t]
\includegraphics[width=0.98\textwidth]{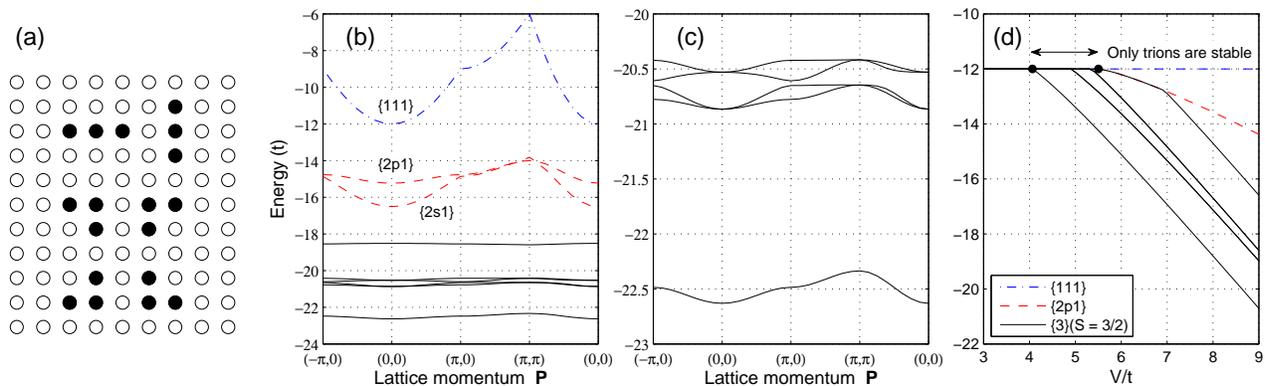}
\caption{(Color online.) Total spin $S = 3/2$. (a) Six basic particle configurations in the strong and medium coupling regimes. (b) A typical six-band spectrum of $S = 3/2$ trions for $V = 10 \, t$ (solid lines). Also shown are the lowest three free fermion energy $E_{\{ 111 \} } = -6 \, t \cos{(P_x/3)} - 6 \, t \cos{(P_y/3)}$ (dotted-dashed line), and the lowest energies of one pair plus one free fermion (dashed lines) for the same total ${\bf P}$. (c) The same trion spectrum at a larger scale. Only five lowest bands are shown. (d) ${\bf P} = (0,0)$ trion energies compared with the lowest $\{ 2p \, 1 \}$ energies. The two middle trion states are double degenerate.}
\label{3UV2d:fig:one}
\end{figure*}

{\em Method.}--- 
The Schr\"odinger equation for the three-fermion wave function $\psi({\bf k}_1, {\bf k}_2, {\bf k}_3)$ is six-dimensional, so direct solution is not practical. The problem can be reduced to a tractable one by making use of two simplifications. (i) Total momentum ${\bf P} = {\bf k}_1 + {\bf k}_2 + {\bf k}_3$ is conserved. Fixing ${\bf P}$ leaves only two two-dimensional variables. (ii) Since the interaction is of finite radius, the interaction part of the Schr\"odinger equation contains a finite number of integrals 
\begin{equation}
F_{i}({\bf q}) = \frac{1}{N} \sum_{\bf k} f({\bf k}) \psi({\bf k},{\bf q},{\bf P}-{\bf q}-{\bf k}) \: , 
\label{3UV2d:eq:four}
\end{equation}
with different permutations of $\psi$'s arguments and $f({\bf k}) = \cos{(\bf kb)}$, $\sin{(\bf kb)}$, or $1$. Expressing $\psi$ as a linear combination of $F_{i}$ and substituting back in Eq.~(\ref{3UV2d:eq:four}) results in a set of coupled integral equations for $F_{i}({\bf q})$. Thus, one two-variable function $\psi$ is replaced by a finite number of one-variable functions $F_{i}$. The latter is a much more tractable problem as long as the interaction radius is not very large. To solve the integral equations, the Brillouin zone is discretized (in this work into $16^2 = 256$ points), ${\bf k}$ integrals are replaced with finite sums, the entire set is transformed into a matrix equation, and the system's energy $E$ is found via eigenvalue search. More details on this reduction methodology can be found in \cite{Rudin1986,Kornilovitch2013,Mattis1986}. The resulting equations are too long to be presented here; in full form, they are given in the Supplemental Material~\cite{SupplMat}. By comparing $8^2$, $12^2$, and $16^2$ discretizations, numerical errors in $16^2$ energies have been estimated to be $< 10^{-3} \, t$, which is sufficient for determining phase boundaries.   

{\em Total spin $S = 3/2$.}--- 
In the ferromagnetic (maximal spin) state, the coordinate wave function is fully antisymmetric. That leaves only one irreducible permutation of $\psi$'s arguments and, given four nearest lattice neighbors, generates four functions $F_i$. The resulting system of four integral equations contains nearest-neighbor attraction $V$ but not contact repulsion $U$.      

In the large-$V$ limit, the fermions are largely confined to the six basic trion configurations shown in Fig.~\ref{3UV2d:fig:one}(a). In the zero order in $t$, the energy is $E^{(0)} = - 2V$. The four ``corner'' configurations are mixed by first-order hopping events. The problem is isomorphic to one-particle motion on a four-site ring. Therefore, one expects a single ground state with $E^{(1)}_1 = - 2V - 2t$, a doublet with $E^{(1)}_{2,3} = - 2V$, and one more trion state with $E^{(1)}_4 = - 2V + 2t$. The energies of the ``linear'' configurations are $E^{(1)}_{5,6} = - 2V$ in this order. Second-order hopping events hybridize the linear configurations with the central doublet of the corner configurations creating a group of four bands with energies near $-2V$.      

An advantage of the present method is the ability to accurately compute the system's energy at arbitrary total momenta ${\bf P}$. A typical trion spectrum is shown in Fig.~\ref{3UV2d:fig:one}(b). The minimal energies of three free fermions $\{ 111 \}$ and of one bound pair plus one free fermion $\{ 21 \}$ with the same ${\bf P}$ are also shown. One should add that calculation of a minimal $\{ 21 \}$ energy is nontrivial, see Supplemental Material for details. As expected, six trion zones are split into groups of one, four, and one. The complex structure of the central band cluster can be seen in Fig.~\ref{3UV2d:fig:one}(c). Notice degeneracies at the high-symmetry points of the Brillouin zone.   

\begin{figure*}[t]
\includegraphics[width=0.98\textwidth]{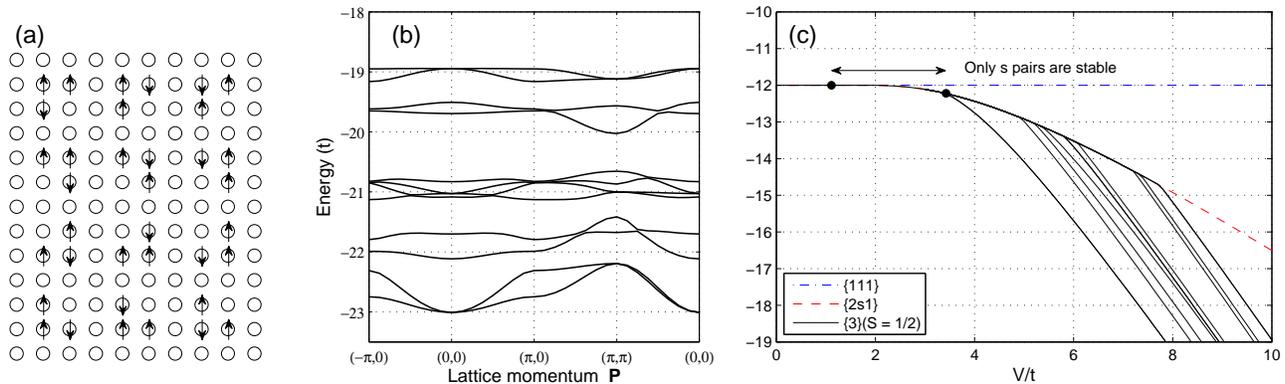}
\caption{(Color online.) Total spin $S = 1/2$. (a) The twelve basic ``corner'' $S_z = 1/2$ configurations of the strong coupling regime. (b) The trion spectrum for $V = U = 10 \, t$. (c) ${\bf P} = (0,0)$ trion energies vs $V$ for $U = 10 \, t$. The first, fifth, and ninth states are double degenerate. For $\frac{10}{9} < V < 3.425$ singlet $s$ pairs are stable whereas trions are unstable.}
\label{3UV2d:fig:two}
\end{figure*}

As $V$ decreases, trion zones begin to disappear into the $\{ 2p \, 1 \}$ continuum. The process proceeds nonuniformly in ${\bf P}$. Trions with small momenta decay first, followed by intermediate and large momenta. Trions with ${\bf P} = (\pi,\pi)$ remain stable to zero $V$. High stability of bound complexes at large lattice momenta is by now well established~\cite{daVeiga2002,Kornilovitch2004,Kornilovitch2013}. Variation of trion energies with $V$ is shown in Fig.~\ref{3UV2d:fig:one}(d). Comparison with $\{ 2p \, 1 \}$ energies reveals a peculiar feature. The triplet pair decays into two free fermions $\{ 11 \}$ at $V = 5.50 \, t$, in accordance with Eq.~(\ref{3UV2d:eq:three}), whereas the $S = 3/2$ trion remains stable until $V = 4.05 \, t$. Thus, if a physical system were limited to only antisymmetric wave functions (as in the case of, e.g., spinless fermions or fully polarized spin-$\frac{1}{2}$ fermions), in the interval $4.05 \, t < V < 5.50 \, t$, trions would be stable while pairs would be unstable. Such an $S = 3/2$ trion is Borromean: all of its two-body subsystems are unstable~\cite{Zhukov1993,Volosniev2013}. This effect might be related to the recently proposed super-Efimov states~\cite{Nishida2013}. However, in the present case the number of trions is limited by the interaction radius and is always finite.

{\em Total spin $S = 1/2$.}--- 
This is a more complex case. In accordance with general rules, wave functions can be chosen antisymmetric with respect to permutation of only two particle momenta, for example ${\bf k}_1$ and ${\bf k}_2$, while not specifying other symmetries~\cite{Landau1977}. This leads to two irreducible arrangements of $\psi$'s arguments in Eq.~(\ref{3UV2d:eq:four}): $\psi({\bf k},{\bf q};{\bf P}-{\bf q}-{\bf k})$ and $\psi({\bf k},{\bf P}-{\bf q}-{\bf k};{\bf q})$. The first argument set generates five functions $F_{i}$, whereas the second generates only four. In the end, the Schr\"odinger equation is reduced to nine coupled two-dimensional integral equations and then converted to $9 \times 16^2 = 2304$ nonsparse linear equations. Since the symmetries involving a third momentum are not constrained, the solutions will include fully antisymmetric ones as a subset. By comparing all the solutions with those obtained within the $S = 3/2$ sector, it is possible to isolate only states with $S = 1/2$. 

Figure~\ref{3UV2d:fig:two}(a) shows the twelve basic corner $S_z = \frac{1}{2}$ configurations of the strong coupling limit. They are mixed by first order hopping within the elementary four-site plaquette. The problem is isomorphic to one particle on a twelve-site ring. In addition, there are six linear configurations (one vertical plus one horizontal times three different positions of the down spin) that are mixed with the corners by second order hopping. Therefore, one expects eighteen trion bands, of which six will be a repeat of the fully antisymmetric $S = 3/2$ bands. By excluding the latter, twelve $S = 1/2$ bands are identified. An exemplary spectrum is shown in Fig.~\ref{3UV2d:fig:two}(b). Comparison with Fig.~\ref{3UV2d:fig:one}(c) shows that the ground state, at ${\bf P} = (0,0)$, belongs to $S = 1/2$ for these $U$ and $V$.

\begin{figure}[b]
\includegraphics[width=0.45\textwidth]{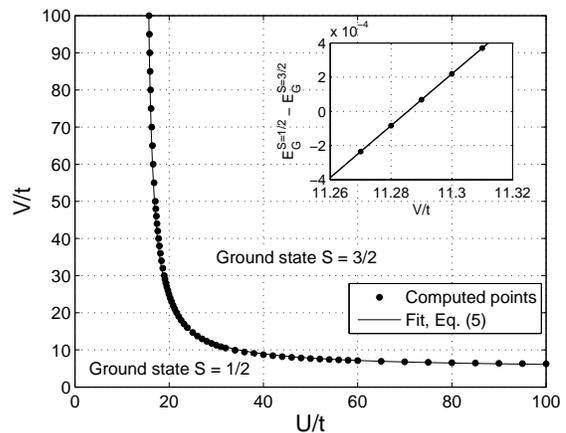}
\caption{Main panel: the Nagaoka boundary line for ${\bf P} = (0,0)$. The solid line is Eq.~(\ref{3UV2d:eq:five}). Inset: the difference between the lowest $S = 1/2$ energy and the lowest $S = 3/2$ energy as a function of $V$ for $U = 30 \, t$. The zero crossing at $V/t = 11.285$ indicates a change in the ground state spin from $S = 1/2$ to $S = 3/2$.}
\label{3UV2d:fig:three}
\end{figure}

Variation of ${\bf P} = (0,0)$ trion energies with $V$ is shown in Fig.~\ref{3UV2d:fig:two}(c). Notice that all the trion states decay into $\{ 2s \, 1 \}$ before the singlet pair dissociates into $\{ 11 \}$. There is a wide $V$ interval where pairs are stable while trions are unstable. When a third fermion is added to an $s$ pair, the wave function must create a node which is analogous of placing the third particle in the first excited state of the twelve-site ring problem. Because of the cluster's finite size, it costs a finite energy (of order $t$), and the trion does not form. At larger $V$, energy gain from forming a second attractive bond exceeds the energy loss driven by the exclusion principle, and the trion forms. The pair stability region observed here is much wider than in the one-dimensional $UV$ model~\cite{Kornilovitch2013}. In 1D, forming a node is equivalent to making $U$ very large. Therefore, the large-$U$ pair and the large-$U$ trion bind almost at the same $V$. In 2D, the node can exist along the angular coordinate, which is not equivalent to a large $U$. The greater stability of pairs against adding a third fermion suggests a finite parameter region with no phase separation in many-body versions of the $UV$ model. This is a welcome result for the local pair mechanisms of superconductivity~\cite{Micnas1990,Alexandrov1994,Alexandrov2010}.       

\begin{figure}[t]
\includegraphics[width=0.45\textwidth]{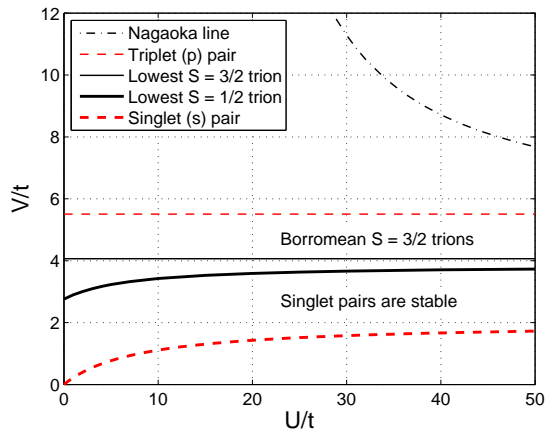}
\caption{(Color online.) Phase boundaries of two-fermion states [for ${\bf K} = (0,0)$] and three-fermion states [for ${\bf P} = (0,0)$] in the two-dimensional $UV$ model. Bound pairs and trions form above the respective lines. The singlet pair line is $V_s$ from Eq.~(\ref{3UV2d:eq:two}). The triplet pair line is $V_p$ from Eq.~(\ref{3UV2d:eq:three}). The $S = 3/2$ line is $V = 4.05 \, t$ (independent of $U$). The Nagaoka line is a fragment of Fig.~\ref{3UV2d:fig:three}.} 
\label{3UV2d:fig:four}
\end{figure}

{\em Nagaoka transition.}--- 
Consider the twelve configurations of Fig.~\ref{3UV2d:fig:two}(a) in the strong coupling limit $U,V \gg t$. The particles are confined to four sites with no double occupancy. The four-site plaquette can, therefore, be thought of as a segment of the two-dimensional half-filled Hubbard model with exactly one hole. Then Nagaoka's arguments apply and a ferromagnetic ground state $S = 3/2$ is expected. For three fermions, Nagaoka's result can also be proven as follows. To first order in $t$, the twelve-site ring problem has a single ground state $\vert \psi_{1} \rangle$ with $E_1 = -2V - 2t$ and a double-degenerate first excited state $\vert \psi_{2,3} \rangle$ with $E_{2,3} = - 2V - \sqrt{3} \, t$. By comparing with the four-site ring of $S = 3/2$, cf. Fig.~\ref{3UV2d:fig:one}(a), one concludes that $\vert \psi_{2,3} \rangle$ belong to the $S = 1/2$ sector while $\vert \psi_{1} \rangle$ may belong to either $S = 3/2$ or $S = 1/2$. Then, it is a simple application of Young diagrams to show that $\vert \psi_{1} \rangle$ cannot have $S = 1/2$ because symmetrization of $\vert \psi_{1} \rangle$ by the two-row diagram yields zero. Thus, the ground state has energy $E_{1}$ and spin $S = 3/2$. At weak coupling, $V \sim t$, the particles are no longer confined within the four sites and the above arguments are not valid. As a result, the ground state should have the lowest spin $S = 1/2$. The plots of Fig.~\ref{3UV2d:fig:two} correspond to this regime. Therefore, one expects an $(S = 1/2) \rightarrow (S = 3/2)$ transition at intermediate $U$ and $V$. A ferromagnetic transition has indeed been found in the exact numerical solution by comparing the trion energies of the $S = 1/2$ and $S = 3/2$ sectors. The boundary line $V_N$ for ${\bf P} = (0,0)$ is shown in Fig.~\ref{3UV2d:fig:three}. It is well approximated by  
\begin{equation}
V_N = \frac{103 \, t^2}{U - 14.6 \, t} + 4.9 \, t  \: . 
\label{3UV2d:eq:five}
\end{equation}

{\em Summary.}--- 
By repeating described procedures at multiple $U$ and $V$, all the phase boundaries have been computed. A composite phase diagram of two-fermion and three-fermion states in the two-dimensional $UV$ model is shown in Fig.~\ref{3UV2d:fig:four}. Three features are of note. A wide area of about $\triangle V \approx 2 \,t$ between the singlet pair and first $S = 1/2$ trion suggests stability of bound pairs against agglomeration and possible absence of phase separation at finite fermion fillings, with implications for superconductivity~\cite{Micnas1990,Alexandrov1994,Alexandrov2010}. Conversely, in the case of fully polarized (or spinless) fermions, trions are more stable than pairs indicating an unusual property of Borromean binding in two dimensions~\cite{Zhukov1993,Volosniev2013}. Finally, in the high-$U$, high-$V$ limit, the three fermion system undergoes a Nagaoka transition into a ferromagnetic state with the total spin $S = 3/2$. The latter two effects may be observable in optical lattice experiments~\cite{Jaksch2005,Bloch2008}.

\begin{acknowledgments}

The author wishes to thank Mona Berciu, Vladimir Bulatov, James Hague, and Jesper Levinsen for helpful discussions on the subject of this Letter.    

\end{acknowledgments}

%\begin{comment}

% Start Supplemental Material

\vspace{1.0cm}

\begin{widetext}

\section{\label{3UV2d:sec:app:a}
Supplemental Material
}

{\em Spin-{\rm 0} bosons}. Although the main focus of the present work is spin-$\frac{1}{2}$ fermions, the equations for spin-$0$ bosons are given below as a useful reference. The wave function is symmetric with respect to permutations of all three arguments. The five reduction functions $F^{0}_1$, $F^{0}_4$, $F^{0}_5$, $F^{0}_8$, $F^{0}_9$ are defined as 
\begin{eqnarray}
F^{0}_1({\bf q}) & = & N^{-1} \sum_{\bf k}             \,        
   \psi^{0}({\bf k},{\bf q},{\bf P} - {\bf q} - {\bf k}) \: ,
\label{3UV2d:eq:oneone}    \\
F^{0}_4({\bf q}) & = & N^{-1} \sum_{\bf k} \cos{(k_x)} \,       
   \psi^{0}({\bf k},{\bf q},{\bf P} - {\bf q} - {\bf k}) \: ,
\label{3UV2d:eq:onetwo}    \\   
F^{0}_5({\bf q}) & = & N^{-1} \sum_{\bf k} \cos{(k_y)} \,       
   \psi^{0}({\bf k},{\bf q},{\bf P} - {\bf q} - {\bf k}) \: ,
\label{3UV2d:eq:onethree}  \\
F^{0}_8({\bf q}) & = & N^{-1} \sum_{\bf k} \sin{(k_x)} \,       
   \psi^{0}({\bf k},{\bf q},{\bf P} - {\bf q} - {\bf k}) \: ,
\label{3UV2d:eq:onefour}   \\
F^{0}_9({\bf q}) & = & N^{-1} \sum_{\bf k} \sin{(k_y)} \,       
   \psi^{0}({\bf k},{\bf q},{\bf P} - {\bf q} - {\bf k}) \: ,
\label{3UV2d:eq:onefive}
\end{eqnarray}
where $N$ is the number of lattice sites. The eigenvalue set of equations is Eqs.~(\ref{3UV2d:eq:onesix})-(\ref{3UV2d:eq:oneten}):
\begin{eqnarray}
F^{0}_1({\bf q}) & = & \frac{U}{N} \sum_{\bf k} 
\frac{F^{0}_1({\bf q}) + 2 F^{0}_1({\bf k})}
{E - \varepsilon({\bf k}) - \varepsilon({\bf q}) - \varepsilon({\bf P}-{\bf q}-{\bf k})}
\nonumber \\
       &   & - \frac{V}{N} \sum_{\bf k}
\frac{[\cos(k_x) + \cos(P_x - q_x - k_x)] F^{0}_4({\bf q}) + 
    2 [\cos(q_x) + \cos(P_x - q_x - k_x)] F^{0}_4({\bf k}) }
{E - \varepsilon({\bf k}) - \varepsilon({\bf q}) - \varepsilon({\bf P}-{\bf q}-{\bf k})}    
\nonumber \\
       &   & - \frac{V}{N} \sum_{\bf k}
\frac{[\cos(k_y) + \cos(P_y - q_y - k_y)] F^{0}_5({\bf q}) + 
    2 [\cos(q_y) + \cos(P_y - q_y - k_y)] F^{0}_5({\bf k}) }
{E - \varepsilon({\bf k}) - \varepsilon({\bf q}) - \varepsilon({\bf P}-{\bf q}-{\bf k})}    
\nonumber \\
       &   & - \frac{V}{N} \sum_{\bf k}
\frac{[\sin(k_x) + \sin(P_x - q_x - k_x)] F^{0}_8({\bf q}) + 
    2 [\sin(q_x) + \sin(P_x - q_x - k_x)] F^{0}_8({\bf k}) }
{E - \varepsilon({\bf k}) - \varepsilon({\bf q}) - \varepsilon({\bf P}-{\bf q}-{\bf k})}    
\nonumber \\
       &   & - \frac{V}{N} \sum_{\bf k}
\frac{[\sin(k_y) + \sin(P_y - q_y - k_y)] F^{0}_9({\bf q}) + 
    2 [\sin(q_y) + \sin(P_y - q_y - k_y)] F^{0}_9({\bf k}) }
{E - \varepsilon({\bf k}) - \varepsilon({\bf q}) - \varepsilon({\bf P}-{\bf q}-{\bf k})} \: ,  
\label{3UV2d:eq:onesix} 
\end{eqnarray}
\begin{eqnarray}
F^{0}_4({\bf q}) & = & \frac{U}{N} \sum_{\bf k} 
\frac{\cos{(k_x)}                          F^{0}_1({\bf q}) +  
     [\cos{(k_x)} + \cos(P_x - q_x - k_x)] F^{0}_1({\bf k})}
{E - \varepsilon({\bf k}) - \varepsilon({\bf q}) - \varepsilon({\bf P}-{\bf q}-{\bf k})}
\nonumber \\
       &   & - \frac{V}{N} \sum_{\bf k}
\frac{\cos{(k_x)} [\cos(k_x) + \cos(P_x - q_x - k_x)] F^{0}_4({\bf q})}
{E - \varepsilon({\bf k}) - \varepsilon({\bf q}) - \varepsilon({\bf P}-{\bf q}-{\bf k})}    
\nonumber \\
       &   & - \frac{V}{N} \sum_{\bf k}
\frac{[\cos(q_x) + \cos(P_x - q_x - k_x)] [\cos(k_x) + \cos(P_x - q_x - k_x)] F^{0}_4({\bf k}) }
{E - \varepsilon({\bf k}) - \varepsilon({\bf q}) - \varepsilon({\bf P}-{\bf q}-{\bf k})}    
\nonumber \\
       &   & - \frac{V}{N} \sum_{\bf k}
\frac{\cos{(k_x)} [\cos(k_y) + \cos(P_y - q_y - k_y)] F^{0}_5({\bf q})}
{E - \varepsilon({\bf k}) - \varepsilon({\bf q}) - \varepsilon({\bf P}-{\bf q}-{\bf k})}    
\nonumber \\
       &   & - \frac{V}{N} \sum_{\bf k}
\frac{[\cos(q_y) + \cos(P_y - q_y - k_y)] [\cos(k_x) + \cos(P_x - q_x - k_x)] F^{0}_5({\bf k}) }
{E - \varepsilon({\bf k}) - \varepsilon({\bf q}) - \varepsilon({\bf P}-{\bf q}-{\bf k})}    
\nonumber \\
       &   & - \frac{V}{N} \sum_{\bf k}
\frac{\cos{(k_x)} [\sin(k_x) + \sin(P_x - q_x - k_x)] F^{0}_8({\bf q})}
{E - \varepsilon({\bf k}) - \varepsilon({\bf q}) - \varepsilon({\bf P}-{\bf q}-{\bf k})}    
\nonumber \\
       &   & - \frac{V}{N} \sum_{\bf k}
\frac{[\sin(q_x) + \sin(P_x - q_x - k_x)] [\cos(k_x) + \cos(P_x - q_x - k_x)] F^{0}_8({\bf k}) }
{E - \varepsilon({\bf k}) - \varepsilon({\bf q}) - \varepsilon({\bf P}-{\bf q}-{\bf k})}    
\nonumber \\
       &   & - \frac{V}{N} \sum_{\bf k}
\frac{\cos{(k_x)} [\sin(k_y) + \sin(P_y - q_y - k_y)] F^{0}_9({\bf q})}
{E - \varepsilon({\bf k}) - \varepsilon({\bf q}) - \varepsilon({\bf P}-{\bf q}-{\bf k})}    
\nonumber \\
       &   & - \frac{V}{N} \sum_{\bf k}
\frac{[\sin(q_y) + \sin(P_y - q_y - k_y)] [\cos(k_x) + \cos(P_x - q_x - k_x)] F^{0}_9({\bf k}) }
{E - \varepsilon({\bf k}) - \varepsilon({\bf q}) - \varepsilon({\bf P}-{\bf q}-{\bf k})}    \: ,
\label{3UV2d:eq:oneseven}
\end{eqnarray}
\begin{eqnarray}
F^{0}_5({\bf q}) & = & \frac{U}{N} \sum_{\bf k} 
\frac{\cos{(k_y)}                          F^{0}_1({\bf q}) +  
     [\cos{(k_y)} + \cos(P_y - q_y - k_y)] F^{0}_1({\bf k})}
{E - \varepsilon({\bf k}) - \varepsilon({\bf q}) - \varepsilon({\bf P}-{\bf q}-{\bf k})}
\nonumber \\
       &   & - \frac{V}{N} \sum_{\bf k}
\frac{\cos{(k_y)} [\cos(k_x) + \cos(P_x - q_x - k_x)] F^{0}_4({\bf q})}
{E - \varepsilon({\bf k}) - \varepsilon({\bf q}) - \varepsilon({\bf P}-{\bf q}-{\bf k})}    
\nonumber \\
       &   & - \frac{V}{N} \sum_{\bf k}
\frac{[\cos(q_x) + \cos(P_x - q_x - k_x)] [\cos(k_y) + \cos(P_y - q_y - k_y)] F^{0}_4({\bf k}) }
{E - \varepsilon({\bf k}) - \varepsilon({\bf q}) - \varepsilon({\bf P}-{\bf q}-{\bf k})}    
\nonumber \\
       &   & - \frac{V}{N} \sum_{\bf k}
\frac{\cos{(k_y)} [\cos(k_y) + \cos(P_y - q_y - k_y)] F^{0}_5({\bf q})}
{E - \varepsilon({\bf k}) - \varepsilon({\bf q}) - \varepsilon({\bf P}-{\bf q}-{\bf k})}    
\nonumber \\
       &   & - \frac{V}{N} \sum_{\bf k}
\frac{[\cos(q_y) + \cos(P_y - q_y - k_y)] [\cos(k_y) + \cos(P_y - q_y - k_y)] F^{0}_5({\bf k}) }
{E - \varepsilon({\bf k}) - \varepsilon({\bf q}) - \varepsilon({\bf P}-{\bf q}-{\bf k})}    
\nonumber \\
       &   & - \frac{V}{N} \sum_{\bf k}
\frac{\cos{(k_y)} [\sin(k_x) + \sin(P_x - q_x - k_x)] F^{0}_8({\bf q})}
{E - \varepsilon({\bf k}) - \varepsilon({\bf q}) - \varepsilon({\bf P}-{\bf q}-{\bf k})}    
\nonumber \\
       &   & - \frac{V}{N} \sum_{\bf k}
\frac{[\sin(q_x) + \sin(P_x - q_x - k_x)] [\cos(k_y) + \cos(P_y - q_y - k_y)] F^{0}_8({\bf k}) }
{E - \varepsilon({\bf k}) - \varepsilon({\bf q}) - \varepsilon({\bf P}-{\bf q}-{\bf k})}    
\nonumber \\
       &   & - \frac{V}{N} \sum_{\bf k}
\frac{\cos{(k_y)} [\sin(k_y) + \sin(P_y - q_y - k_y)] F^{0}_9({\bf q})}
{E - \varepsilon({\bf k}) - \varepsilon({\bf q}) - \varepsilon({\bf P}-{\bf q}-{\bf k})}    
\nonumber \\
       &   & - \frac{V}{N} \sum_{\bf k}
\frac{[\sin(q_y) + \sin(P_y - q_y - k_y)] [\cos(k_y) + \cos(P_y - q_y - k_y)] F^{0}_9({\bf k}) }
{E - \varepsilon({\bf k}) - \varepsilon({\bf q}) - \varepsilon({\bf P}-{\bf q}-{\bf k})}    \: ,
\label{3UV2d:eq:oneeight}
\end{eqnarray}
\begin{eqnarray}
F^{0}_8({\bf q}) & = & \frac{U}{N} \sum_{\bf k} 
\frac{\sin{(k_x)}                          F^{0}_1({\bf q}) +  
     [\sin{(k_x)} + \sin(P_x - q_x - k_x)] F^{0}_1({\bf k})}
{E - \varepsilon({\bf k}) - \varepsilon({\bf q}) - \varepsilon({\bf P}-{\bf q}-{\bf k})}
\nonumber \\
       &   & - \frac{V}{N} \sum_{\bf k}
\frac{\sin{(k_x)} [\cos(k_x) + \cos(P_x - q_x - k_x)] F^{0}_4({\bf q})}
{E - \varepsilon({\bf k}) - \varepsilon({\bf q}) - \varepsilon({\bf P}-{\bf q}-{\bf k})}    
\nonumber \\
       &   & - \frac{V}{N} \sum_{\bf k}
\frac{[\cos(q_x) + \cos(P_x - q_x - k_x)] [\sin(k_x) + \sin(P_x - q_x - k_x)] F^{0}_4({\bf k}) }
{E - \varepsilon({\bf k}) - \varepsilon({\bf q}) - \varepsilon({\bf P}-{\bf q}-{\bf k})}    
\nonumber \\
       &   & - \frac{V}{N} \sum_{\bf k}
\frac{\sin{(k_x)} [\cos(k_y) + \cos(P_y - q_y - k_y)] F^{0}_5({\bf q})}
{E - \varepsilon({\bf k}) - \varepsilon({\bf q}) - \varepsilon({\bf P}-{\bf q}-{\bf k})}    
\nonumber \\
       &   & - \frac{V}{N} \sum_{\bf k}
\frac{[\cos(q_y) + \cos(P_y - q_y - k_y)] [\sin(k_x) + \sin(P_x - q_x - k_x)] F^{0}_5({\bf k}) }
{E - \varepsilon({\bf k}) - \varepsilon({\bf q}) - \varepsilon({\bf P}-{\bf q}-{\bf k})}    
\nonumber \\
       &   & - \frac{V}{N} \sum_{\bf k}
\frac{\sin{(k_x)} [\sin(k_x) + \sin(P_x - q_x - k_x)] F^{0}_8({\bf q})}
{E - \varepsilon({\bf k}) - \varepsilon({\bf q}) - \varepsilon({\bf P}-{\bf q}-{\bf k})}    
\nonumber \\
       &   & - \frac{V}{N} \sum_{\bf k}
\frac{[\sin(q_x) + \sin(P_x - q_x - k_x)] [\sin(k_x) + \sin(P_x - q_x - k_x)] F^{0}_8({\bf k}) }
{E - \varepsilon({\bf k}) - \varepsilon({\bf q}) - \varepsilon({\bf P}-{\bf q}-{\bf k})}    
\nonumber \\
       &   & - \frac{V}{N} \sum_{\bf k}
\frac{\sin{(k_x)} [\sin(k_y) + \sin(P_y - q_y - k_y)] F^{0}_9({\bf q})}
{E - \varepsilon({\bf k}) - \varepsilon({\bf q}) - \varepsilon({\bf P}-{\bf q}-{\bf k})}    
\nonumber \\
       &   & - \frac{V}{N} \sum_{\bf k}
\frac{[\sin(q_y) + \sin(P_y - q_y - k_y)] [\sin(k_x) + \sin(P_x - q_x - k_x)] F^{0}_9({\bf k}) }
{E - \varepsilon({\bf k}) - \varepsilon({\bf q}) - \varepsilon({\bf P}-{\bf q}-{\bf k})}    \: ,
\label{3UV2d:eq:onenine}
\end{eqnarray}
\begin{eqnarray}
F^{0}_9({\bf q}) & = & \frac{U}{N} \sum_{\bf k} 
\frac{\sin{(k_y)}                          F^{0}_1({\bf q}) +  
     [\sin{(k_y)} + \sin(P_y - q_y - k_y)] F^{0}_1({\bf k})}
{E - \varepsilon({\bf k}) - \varepsilon({\bf q}) - \varepsilon({\bf P}-{\bf q}-{\bf k})}
\nonumber \\
       &   & - \frac{V}{N} \sum_{\bf k}
\frac{\sin{(k_y)} [\cos(k_x) + \cos(P_x - q_x - k_x)] F^{0}_4({\bf q})}
{E - \varepsilon({\bf k}) - \varepsilon({\bf q}) - \varepsilon({\bf P}-{\bf q}-{\bf k})}    
\nonumber \\
       &   & - \frac{V}{N} \sum_{\bf k}
\frac{[\cos(q_x) + \cos(P_x - q_x - k_x)] [\sin(k_y) + \sin(P_y - q_y - k_y)] F^{0}_4({\bf k}) }
{E - \varepsilon({\bf k}) - \varepsilon({\bf q}) - \varepsilon({\bf P}-{\bf q}-{\bf k})}    
\nonumber \\
       &   & - \frac{V}{N} \sum_{\bf k}
\frac{\sin{(k_y)} [\cos(k_y) + \cos(P_y - q_y - k_y)] F^{0}_5({\bf q})}
{E - \varepsilon({\bf k}) - \varepsilon({\bf q}) - \varepsilon({\bf P}-{\bf q}-{\bf k})}    
\nonumber \\
       &   & - \frac{V}{N} \sum_{\bf k}
\frac{[\cos(q_y) + \cos(P_y - q_y - k_y)] [\sin(k_y) + \sin(P_y - q_y - k_y)] F^{0}_5({\bf k}) }
{E - \varepsilon({\bf k}) - \varepsilon({\bf q}) - \varepsilon({\bf P}-{\bf q}-{\bf k})}    
\nonumber \\
       &   & - \frac{V}{N} \sum_{\bf k}
\frac{\sin{(k_y)} [\sin(k_x) + \sin(P_x - q_x - k_x)] F^{0}_8({\bf q})}
{E - \varepsilon({\bf k}) - \varepsilon({\bf q}) - \varepsilon({\bf P}-{\bf q}-{\bf k})}    
\nonumber \\
       &   & - \frac{V}{N} \sum_{\bf k}
\frac{[\sin(q_x) + \sin(P_x - q_x - k_x)] [\sin(k_y) + \sin(P_y - q_y - k_y)] F^{0}_8({\bf k}) }
{E - \varepsilon({\bf k}) - \varepsilon({\bf q}) - \varepsilon({\bf P}-{\bf q}-{\bf k})}    
\nonumber \\
       &   & - \frac{V}{N} \sum_{\bf k}
\frac{\sin{(k_y)} [\sin(k_y) + \sin(P_y - q_y - k_y)] F^{0}_9({\bf q})}
{E - \varepsilon({\bf k}) - \varepsilon({\bf q}) - \varepsilon({\bf P}-{\bf q}-{\bf k})}    
\nonumber \\
       &   & - \frac{V}{N} \sum_{\bf k}
\frac{[\sin(q_y) + \sin(P_y - q_y - k_y)] [\sin(k_y) + \sin(P_y - q_y - k_y)] F^{0}_9({\bf k}) }
{E - \varepsilon({\bf k}) - \varepsilon({\bf q}) - \varepsilon({\bf P}-{\bf q}-{\bf k})}    \: ,
\label{3UV2d:eq:oneten}
\end{eqnarray}
where the one-particle spectrum $\varepsilon({\bf k})$ is
\begin{equation}
\varepsilon({\bf k}) = - 2t \cos{(k_x)} - 2t \cos{(k_y)} \: . 
\label{3UV2d:eq:onetenone}
\end{equation}

{\em Total spin} $S = 3/2$. The wave function $\psi$ is antisymmetric with respect to permutation of all three arguments. The four reduction functions $F^{3/2}_4$, $F^{3/2}_5$, $F^{3/2}_8$, $F^{3/2}_9$ are defined as 
\begin{eqnarray}
F^{3/2}_4({\bf q}) & = & N^{-1} \sum_{\bf k} \cos{(k_x)} \,       
   \psi^{3/2}({\bf k},{\bf q},{\bf P} - {\bf q} - {\bf k}) \: ,
\label{3UV2d:eq:oneeleven}    \\   
F^{3/2}_5({\bf q}) & = & N^{-1} \sum_{\bf k} \cos{(k_y)} \,       
   \psi^{3/2}({\bf k},{\bf q},{\bf P} - {\bf q} - {\bf k}) \: ,
\label{3UV2d:eq:onetwelve}  \\
F^{3/2}_8({\bf q}) & = & N^{-1} \sum_{\bf k} \sin{(k_x)} \,       
   \psi^{3/2}({\bf k},{\bf q},{\bf P} - {\bf q} - {\bf k}) \: ,
\label{3UV2d:eq:onethirteen}   \\
F^{3/2}_9({\bf q}) & = & N^{-1} \sum_{\bf k} \sin{(k_y)} \,       
   \psi^{3/2}({\bf k},{\bf q},{\bf P} - {\bf q} - {\bf k}) \: .
\label{3UV2d:eq:onefourteen}
\end{eqnarray}
Note that due to the antisymmetry of $\psi$, a function $F^{3/2}({\bf q}) = N^{-1} \sum_{\bf k} \psi^{3/2}({\bf k},{\bf q},{\bf P} - {\bf q} - {\bf k})$ corresponding to on-site interaction vanishes identically and does not appear in the full system. For the same reason, the resulting equations do not include on-site interaction potential $U$. The eigenvalue equations are Eqs.~(\ref{3UV2d:eq:onefifteen})-(\ref{3UV2d:eq:oneeighteen}).  
\begin{eqnarray}
F^{3/2}_4({\bf q}) & = & - \frac{V}{N} \sum_{\bf k}
\frac{\cos{(k_x)} [\cos(k_x) - \cos(P_x - q_x - k_x)] F^{3/2}_4({\bf q})}
{E - \varepsilon({\bf k}) - \varepsilon({\bf q}) - \varepsilon({\bf P}-{\bf q}-{\bf k})}    
\nonumber \\
       &   & - \frac{V}{N} \sum_{\bf k}
\frac{[\cos(q_x) - \cos(P_x - q_x - k_x)] [\cos(P_x - q_x - k_x) - \cos(k_x)] F^{3/2}_4({\bf k}) }
{E - \varepsilon({\bf k}) - \varepsilon({\bf q}) - \varepsilon({\bf P}-{\bf q}-{\bf k})}    
\nonumber \\
       &   & - \frac{V}{N} \sum_{\bf k}
\frac{\cos{(k_x)} [\cos(k_y) - \cos(P_y - q_y - k_y)] F^{3/2}_5({\bf q})}
{E - \varepsilon({\bf k}) - \varepsilon({\bf q}) - \varepsilon({\bf P}-{\bf q}-{\bf k})}    
\nonumber \\
       &   & - \frac{V}{N} \sum_{\bf k}
\frac{[\cos(q_y) - \cos(P_y - q_y - k_y)] [\cos(P_x - q_x - k_x) - \cos(k_x)] F^{3/2}_5({\bf k}) }
{E - \varepsilon({\bf k}) - \varepsilon({\bf q}) - \varepsilon({\bf P}-{\bf q}-{\bf k})}    
\nonumber \\
       &   & - \frac{V}{N} \sum_{\bf k}
\frac{\cos{(k_x)} [\sin(k_x) - \sin(P_x - q_x - k_x)] F^{3/2}_8({\bf q})}
{E - \varepsilon({\bf k}) - \varepsilon({\bf q}) - \varepsilon({\bf P}-{\bf q}-{\bf k})}    
\nonumber \\
       &   & - \frac{V}{N} \sum_{\bf k}
\frac{[\sin(q_x) - \sin(P_x - q_x - k_x)] [\cos(P_x - q_x - k_x) - \cos(k_x)] F^{3/2}_8({\bf k}) }
{E - \varepsilon({\bf k}) - \varepsilon({\bf q}) - \varepsilon({\bf P}-{\bf q}-{\bf k})}    
\nonumber \\
       &   & - \frac{V}{N} \sum_{\bf k}
\frac{\cos{(k_x)} [\sin(k_y) - \sin(P_y - q_y - k_y)] F^{3/2}_9({\bf q})}
{E - \varepsilon({\bf k}) - \varepsilon({\bf q}) - \varepsilon({\bf P}-{\bf q}-{\bf k})}    
\nonumber \\
       &   & - \frac{V}{N} \sum_{\bf k}
\frac{[\sin(q_y) - \sin(P_y - q_y - k_y)] [\cos(P_x - q_x - k_x) - \cos(k_x)] F^{3/2}_9({\bf k}) }
{E - \varepsilon({\bf k}) - \varepsilon({\bf q}) - \varepsilon({\bf P}-{\bf q}-{\bf k})}    \: ,
\label{3UV2d:eq:onefifteen}
\end{eqnarray}
\begin{eqnarray}
F^{3/2}_5({\bf q}) & = & - \frac{V}{N} \sum_{\bf k}
\frac{\cos{(k_y)} [\cos(k_x) - \cos(P_x - q_x - k_x)] F^{3/2}_4({\bf q})}
{E - \varepsilon({\bf k}) - \varepsilon({\bf q}) - \varepsilon({\bf P}-{\bf q}-{\bf k})}    
\nonumber \\
       &   & - \frac{V}{N} \sum_{\bf k}
\frac{[\cos(q_x) - \cos(P_x - q_x - k_x)] [\cos(P_y - q_y - k_y) - \cos(k_y)] F^{3/2}_4({\bf k}) }
{E - \varepsilon({\bf k}) - \varepsilon({\bf q}) - \varepsilon({\bf P}-{\bf q}-{\bf k})}    
\nonumber \\
       &   & - \frac{V}{N} \sum_{\bf k}
\frac{\cos{(k_y)} [\cos(k_y) - \cos(P_y - q_y - k_y)] F^{3/2}_5({\bf q})}
{E - \varepsilon({\bf k}) - \varepsilon({\bf q}) - \varepsilon({\bf P}-{\bf q}-{\bf k})}    
\nonumber \\
       &   & - \frac{V}{N} \sum_{\bf k}
\frac{[\cos(q_y) - \cos(P_y - q_y - k_y)] [\cos(P_y - q_y - k_y) - \cos(k_y)] F^{3/2}_5({\bf k}) }
{E - \varepsilon({\bf k}) - \varepsilon({\bf q}) - \varepsilon({\bf P}-{\bf q}-{\bf k})}    
\nonumber \\
       &   & - \frac{V}{N} \sum_{\bf k}
\frac{\cos{(k_y)} [\sin(k_x) - \sin(P_x - q_x - k_x)] F^{3/2}_8({\bf q})}
{E - \varepsilon({\bf k}) - \varepsilon({\bf q}) - \varepsilon({\bf P}-{\bf q}-{\bf k})}    
\nonumber \\
       &   & - \frac{V}{N} \sum_{\bf k}
\frac{[\sin(q_x) - \sin(P_x - q_x - k_x)] [\cos(P_y - q_y - k_y) - \cos(k_y)] F^{3/2}_8({\bf k}) }
{E - \varepsilon({\bf k}) - \varepsilon({\bf q}) - \varepsilon({\bf P}-{\bf q}-{\bf k})}    
\nonumber \\
       &   & - \frac{V}{N} \sum_{\bf k}
\frac{\cos{(k_y)} [\sin(k_y) - \sin(P_y - q_y - k_y)] F^{3/2}_9({\bf q})}
{E - \varepsilon({\bf k}) - \varepsilon({\bf q}) - \varepsilon({\bf P}-{\bf q}-{\bf k})}    
\nonumber \\
       &   & - \frac{V}{N} \sum_{\bf k}
\frac{[\sin(q_y) - \sin(P_y - q_y - k_y)] [\cos(P_y - q_y - k_y) - \cos(k_y)] F^{3/2}_9({\bf k}) }
{E - \varepsilon({\bf k}) - \varepsilon({\bf q}) - \varepsilon({\bf P}-{\bf q}-{\bf k})}    \: ,
\label{3UV2d:eq:onesixteen}
\end{eqnarray}
\begin{eqnarray}
F^{3/2}_8({\bf q}) & = & - \frac{V}{N} \sum_{\bf k}
\frac{\sin{(k_x)} [\cos(k_x) - \cos(P_x - q_x - k_x)] F^{3/2}_4({\bf q})}
{E - \varepsilon({\bf k}) - \varepsilon({\bf q}) - \varepsilon({\bf P}-{\bf q}-{\bf k})}    
\nonumber \\
       &   & - \frac{V}{N} \sum_{\bf k}
\frac{[\cos(q_x) - \cos(P_x - q_x - k_x)] [\sin(P_x - q_x - k_x) - \sin(k_x)] F^{3/2}_4({\bf k}) }
{E - \varepsilon({\bf k}) - \varepsilon({\bf q}) - \varepsilon({\bf P}-{\bf q}-{\bf k})}    
\nonumber \\
       &   & - \frac{V}{N} \sum_{\bf k}
\frac{\sin{(k_x)} [\cos(k_y) - \cos(P_y - q_y - k_y)] F^{3/2}_5({\bf q})}
{E - \varepsilon({\bf k}) - \varepsilon({\bf q}) - \varepsilon({\bf P}-{\bf q}-{\bf k})}    
\nonumber \\
       &   & - \frac{V}{N} \sum_{\bf k}
\frac{[\cos(q_y) - \cos(P_y - q_y - k_y)] [\sin(P_x - q_x - k_x) - \sin(k_x)] F^{3/2}_5({\bf k}) }
{E - \varepsilon({\bf k}) - \varepsilon({\bf q}) - \varepsilon({\bf P}-{\bf q}-{\bf k})}    
\nonumber \\
       &   & - \frac{V}{N} \sum_{\bf k}
\frac{\sin{(k_x)} [\sin(k_x) - \sin(P_x - q_x - k_x)] F^{3/2}_8({\bf q})}
{E - \varepsilon({\bf k}) - \varepsilon({\bf q}) - \varepsilon({\bf P}-{\bf q}-{\bf k})}    
\nonumber \\
       &   & - \frac{V}{N} \sum_{\bf k}
\frac{[\sin(q_x) - \sin(P_x - q_x - k_x)] [\sin(P_x - q_x - k_x) - \sin(k_x)] F^{3/2}_8({\bf k}) }
{E - \varepsilon({\bf k}) - \varepsilon({\bf q}) - \varepsilon({\bf P}-{\bf q}-{\bf k})}    
\nonumber \\
       &   & - \frac{V}{N} \sum_{\bf k}
\frac{\sin{(k_x)} [\sin(k_y) - \sin(P_y - q_y - k_y)] F^{3/2}_9({\bf q})}
{E - \varepsilon({\bf k}) - \varepsilon({\bf q}) - \varepsilon({\bf P}-{\bf q}-{\bf k})}    
\nonumber \\
       &   & - \frac{V}{N} \sum_{\bf k}
\frac{[\sin(q_y) - \sin(P_y - q_y - k_y)] [\sin(P_x - q_x - k_x) - \sin(k_x)] F^{3/2}_9({\bf k}) }
{E - \varepsilon({\bf k}) - \varepsilon({\bf q}) - \varepsilon({\bf P}-{\bf q}-{\bf k})}    \: ,
\label{3UV2d:eq:oneseventeen}
\end{eqnarray}
\begin{eqnarray}
F^{3/2}_9({\bf q}) & = & - \frac{V}{N} \sum_{\bf k}
\frac{\sin{(k_y)} [\cos(k_x) - \cos(P_x - q_x - k_x)] F^{3/2}_4({\bf q})}
{E - \varepsilon({\bf k}) - \varepsilon({\bf q}) - \varepsilon({\bf P}-{\bf q}-{\bf k})}    
\nonumber \\
       &   & - \frac{V}{N} \sum_{\bf k}
\frac{[\cos(q_x) - \cos(P_x - q_x - k_x)] [\sin(P_y - q_y - k_y) - \sin(k_y)] F^{3/2}_4({\bf k}) }
{E - \varepsilon({\bf k}) - \varepsilon({\bf q}) - \varepsilon({\bf P}-{\bf q}-{\bf k})}    
\nonumber \\
       &   & - \frac{V}{N} \sum_{\bf k}
\frac{\sin{(k_y)} [\cos(k_y) - \cos(P_y - q_y - k_y)] F^{3/2}_5({\bf q})}
{E - \varepsilon({\bf k}) - \varepsilon({\bf q}) - \varepsilon({\bf P}-{\bf q}-{\bf k})}    
\nonumber \\
       &   & - \frac{V}{N} \sum_{\bf k}
\frac{[\cos(q_y) - \cos(P_y - q_y - k_y)] [\sin(P_y - q_y - k_y) - \sin(k_y)] F^{3/2}_5({\bf k}) }
{E - \varepsilon({\bf k}) - \varepsilon({\bf q}) - \varepsilon({\bf P}-{\bf q}-{\bf k})}    
\nonumber \\
       &   & - \frac{V}{N} \sum_{\bf k}
\frac{\sin{(k_y)} [\sin(k_x) - \sin(P_x - q_x - k_x)] F^{3/2}_8({\bf q})}
{E - \varepsilon({\bf k}) - \varepsilon({\bf q}) - \varepsilon({\bf P}-{\bf q}-{\bf k})}    
\nonumber \\
       &   & - \frac{V}{N} \sum_{\bf k}
\frac{[\sin(q_x) - \sin(P_x - q_x - k_x)] [\sin(P_y - q_y - k_y) - \sin(k_y)] F^{3/2}_8({\bf k}) }
{E - \varepsilon({\bf k}) - \varepsilon({\bf q}) - \varepsilon({\bf P}-{\bf q}-{\bf k})}    
\nonumber \\
       &   & - \frac{V}{N} \sum_{\bf k}
\frac{\sin{(k_y)} [\sin(k_y) - \sin(P_y - q_y - k_y)] F^{3/2}_9({\bf q})}
{E - \varepsilon({\bf k}) - \varepsilon({\bf q}) - \varepsilon({\bf P}-{\bf q}-{\bf k})}    
\nonumber \\
       &   & - \frac{V}{N} \sum_{\bf k}
\frac{[\sin(q_y) - \sin(P_y - q_y - k_y)] [\sin(P_y - q_y - k_y) - \sin(k_y)] F^{3/2}_9({\bf k}) }
{E - \varepsilon({\bf k}) - \varepsilon({\bf q}) - \varepsilon({\bf P}-{\bf q}-{\bf k})}    \: .
\label{3UV2d:eq:oneeighteen}
\end{eqnarray}

{\em Total spin} $S = 1/2$. The wave function is antisymmetric with respect to permutation of the first and second arguments. The nine reduction functions $F^{1/2}_1$, $F^{1/2}_2$, $F^{1/2}_3$, $F^{1/2}_4$, $F^{1/2}_5$, $F^{1/2}_6$, $F^{1/2}_7$, $F^{1/2}_8$, $F^{1/2}_9$ are defined as 
\begin{eqnarray}
F^{1/2}_1({\bf q}) & = & N^{-1} \sum_{\bf k}        
   \psi^{1/2}({\bf k},{\bf q},{\bf P} - {\bf q} - {\bf k}) \: ,
\label{3UV2d:eq:onetwnetyone}    \\   
F^{1/2}_2({\bf q}) & = & N^{-1} \sum_{\bf k} \cos{(k_x)} \,       
   \psi^{1/2}({\bf k},{\bf P} - {\bf q} - {\bf k},{\bf q}) \: ,
\label{3UV2d:eq:onetwentytwo}  \\
F^{1/2}_3({\bf q}) & = & N^{-1} \sum_{\bf k} \cos{(k_y)} \,       
   \psi^{1/2}({\bf k},{\bf P} - {\bf q} - {\bf k},{\bf q}) \: ,
\label{3UV2d:eq:onetwentythree}   \\
F^{1/2}_4({\bf q}) & = & N^{-1} \sum_{\bf k} \cos{(k_x)} \,       
   \psi^{1/2}({\bf k},{\bf q},{\bf P} - {\bf q} - {\bf k}) \: ,
\label{3UV2d:eq:onetwentyfour}    \\   
F^{1/2}_5({\bf q}) & = & N^{-1} \sum_{\bf k} \cos{(k_y)} \,       
   \psi^{1/2}({\bf k},{\bf q},{\bf P} - {\bf q} - {\bf k}) \: ,
\label{3UV2d:eq:onetwentyfive}  \\
F^{1/2}_6({\bf q}) & = & N^{-1} \sum_{\bf k} \sin{(k_x)} \,       
   \psi^{1/2}({\bf k},{\bf P} - {\bf q} - {\bf k},{\bf q}) \: ,
\label{3UV2d:eq:onetwentysix}  \\
F^{1/2}_7({\bf q}) & = & N^{-1} \sum_{\bf k} \sin{(k_y)} \,       
   \psi^{1/2}({\bf k},{\bf P} - {\bf q} - {\bf k},{\bf q}) \: ,
\label{3UV2d:eq:onetwentyseven}   \\
F^{1/2}_8({\bf q}) & = & N^{-1} \sum_{\bf k} \sin{(k_x)} \,       
   \psi^{1/2}({\bf k},{\bf q},{\bf P} - {\bf q} - {\bf k}) \: ,
\label{3UV2d:eq:onetwentyeight}   \\
F^{1/2}_9({\bf q}) & = & N^{-1} \sum_{\bf k} \sin{(k_y)} \,       
   \psi^{1/2}({\bf k},{\bf q},{\bf P} - {\bf q} - {\bf k}) \: .
\label{3UV2d:eq:onetwentynine}
\end{eqnarray}
Similar to the $S = 3/2$ case, here a function $F^{1/2}({\bf q}) = N^{-1} \sum_{\bf k} \psi^{1/2}({\bf k},{\bf P} - {\bf q} - {\bf k},{\bf q})$ vanishes identically and drops out from the set. The eigenvalue equations are Eqs.~(\ref{3UV2d:eq:onethirtyone})-(\ref{3UV2d:eq:onethirtynine}):
\begin{eqnarray}
F^{1/2}_1({\bf q}) & = & \frac{U}{N} \sum_{\bf k} 
\frac{F^{1/2}_1({\bf q}) - F^{1/2}_1({\bf k})}
{E - \varepsilon({\bf k}) - \varepsilon({\bf q}) - \varepsilon({\bf P}-{\bf q}-{\bf k})}
\nonumber \\
       &   & - \frac{V}{N} \sum_{\bf k}
\frac{[\cos(P_x - q_x - k_x) - \cos(q_x)] F^{1/2}_2({\bf k}) }
{E - \varepsilon({\bf k}) - \varepsilon({\bf q}) - \varepsilon({\bf P}-{\bf q}-{\bf k})}    
\nonumber \\
       &   & - \frac{V}{N} \sum_{\bf k}
\frac{[\cos(P_y - q_y - k_y) - \cos(q_y)] F^{1/2}_3({\bf k}) }
{E - \varepsilon({\bf k}) - \varepsilon({\bf q}) - \varepsilon({\bf P}-{\bf q}-{\bf k})}    
\nonumber \\
       &   & - \frac{V}{N} \sum_{\bf k}
\frac{2 \cos(k_x) \, F^{1/2}_4({\bf q}) - 2 \cos(q_x) \, F^{1/2}_4({\bf k}) }
{E - \varepsilon({\bf k}) - \varepsilon({\bf q}) - \varepsilon({\bf P}-{\bf q}-{\bf k})}    
\nonumber \\
       &   & - \frac{V}{N} \sum_{\bf k}
\frac{2 \cos(k_y) \, F^{1/2}_5({\bf q}) - 2 \cos(q_y) \, F^{1/2}_5({\bf k}) }
{E - \varepsilon({\bf k}) - \varepsilon({\bf q}) - \varepsilon({\bf P}-{\bf q}-{\bf k})}    
\nonumber \\
       &   & - \frac{V}{N} \sum_{\bf k}
\frac{[\sin(P_x - q_x - k_x) - \sin(q_x)] F^{1/2}_6({\bf k}) }
{E - \varepsilon({\bf k}) - \varepsilon({\bf q}) - \varepsilon({\bf P}-{\bf q}-{\bf k})}    
\nonumber \\
       &   & - \frac{V}{N} \sum_{\bf k}
\frac{[\sin(P_y - q_y - k_y) - \sin(q_y)] F^{1/2}_7({\bf k}) }
{E - \varepsilon({\bf k}) - \varepsilon({\bf q}) - \varepsilon({\bf P}-{\bf q}-{\bf k})}    
\nonumber \\
       &   & - \frac{V}{N} \sum_{\bf k}
\frac{2 \sin(k_x) \, F^{1/2}_8({\bf q}) - 2 \sin(q_x) \, F^{1/2}_8({\bf k}) }
{E - \varepsilon({\bf k}) - \varepsilon({\bf q}) - \varepsilon({\bf P}-{\bf q}-{\bf k})}    
\nonumber \\
       &   & - \frac{V}{N} \sum_{\bf k}
\frac{2 \sin(k_y) \, F^{1/2}_9({\bf q}) - 2 \sin(q_y) \, F^{1/2}_9({\bf k}) }
{E - \varepsilon({\bf k}) - \varepsilon({\bf q}) - \varepsilon({\bf P}-{\bf q}-{\bf k})}  \: ,  
\label{3UV2d:eq:onethirtyone} 
\end{eqnarray}
\begin{eqnarray}
F^{1/2}_2({\bf q}) & = & \frac{U}{N} \sum_{\bf k} 
\frac{[\cos(P_x - q_x - k_x) - \cos(k_x)] F^{1/2}_1({\bf k})}
{E - \varepsilon({\bf k}) - \varepsilon({\bf q}) - \varepsilon({\bf P}-{\bf q}-{\bf k})}
\nonumber \\
       &   & - \frac{V}{N} \sum_{\bf k}
\frac{\cos(k_x) [\cos(k_x) - \cos(P_x - q_x - k_x)] F^{1/2}_2({\bf q}) }
{E - \varepsilon({\bf k}) - \varepsilon({\bf q}) - \varepsilon({\bf P}-{\bf q}-{\bf k})}    
\nonumber \\
       &   & - \frac{V}{N} \sum_{\bf k}
\frac{\cos(k_x) [\cos(k_y) - \cos(P_y - q_y - k_y)] F^{1/2}_3({\bf q}) }
{E - \varepsilon({\bf k}) - \varepsilon({\bf q}) - \varepsilon({\bf P}-{\bf q}-{\bf k})}    
\nonumber \\
       &   & - \frac{V}{N} \sum_{\bf k}
\frac{2 \cos(P_x - q_x - k_x) [\cos(P_x - q_x - k_x) - \cos(k_x)] F^{1/2}_4({\bf k}) }
{E - \varepsilon({\bf k}) - \varepsilon({\bf q}) - \varepsilon({\bf P}-{\bf q}-{\bf k})}    
\nonumber \\
       &   & - \frac{V}{N} \sum_{\bf k}
\frac{2 \cos(P_y - q_y - k_y) [\cos(P_x - q_x - k_x) - \cos(k_x)] F^{1/2}_5({\bf k}) }
{E - \varepsilon({\bf k}) - \varepsilon({\bf q}) - \varepsilon({\bf P}-{\bf q}-{\bf k})}    
\nonumber \\
       &   & - \frac{V}{N} \sum_{\bf k}
\frac{\cos(k_x) [\sin(k_x) - \sin(P_x - q_x - k_x)] F^{1/2}_6({\bf q}) }
{E - \varepsilon({\bf k}) - \varepsilon({\bf q}) - \varepsilon({\bf P}-{\bf q}-{\bf k})}    
\nonumber \\
       &   & - \frac{V}{N} \sum_{\bf k}
\frac{\cos(k_x) [\sin(k_y) - \sin(P_y - q_y - k_y)] F^{1/2}_7({\bf q}) }
{E - \varepsilon({\bf k}) - \varepsilon({\bf q}) - \varepsilon({\bf P}-{\bf q}-{\bf k})}    
\nonumber \\
       &   & - \frac{V}{N} \sum_{\bf k}
\frac{2 \sin(P_x - q_x - k_x) [\cos(P_x - q_x - k_x) - \cos(k_x)] F^{1/2}_8({\bf k}) }
{E - \varepsilon({\bf k}) - \varepsilon({\bf q}) - \varepsilon({\bf P}-{\bf q}-{\bf k})}    
\nonumber \\
       &   & - \frac{V}{N} \sum_{\bf k}
\frac{2 \sin(P_y - q_y - k_y) [\cos(P_x - q_x - k_x) - \cos(k_x)] F^{1/2}_9({\bf k}) }
{E - \varepsilon({\bf k}) - \varepsilon({\bf q}) - \varepsilon({\bf P}-{\bf q}-{\bf k})}  \: ,  
\label{3UV2d:eq:onethirtytwo} 
\end{eqnarray}
\begin{eqnarray}
F^{1/2}_3({\bf q}) & = & \frac{U}{N} \sum_{\bf k} 
\frac{[\cos(P_y - q_y - k_y) - \cos(k_y)] F^{1/2}_1({\bf k})}
{E - \varepsilon({\bf k}) - \varepsilon({\bf q}) - \varepsilon({\bf P}-{\bf q}-{\bf k})}
\nonumber \\
       &   & - \frac{V}{N} \sum_{\bf k}
\frac{\cos(k_y) [\cos(k_x) - \cos(P_x - q_x - k_x)] F^{1/2}_2({\bf q}) }
{E - \varepsilon({\bf k}) - \varepsilon({\bf q}) - \varepsilon({\bf P}-{\bf q}-{\bf k})}    
\nonumber \\
       &   & - \frac{V}{N} \sum_{\bf k}
\frac{\cos(k_y) [\cos(k_y) - \cos(P_y - q_y - k_y)] F^{1/2}_3({\bf q}) }
{E - \varepsilon({\bf k}) - \varepsilon({\bf q}) - \varepsilon({\bf P}-{\bf q}-{\bf k})}    
\nonumber \\
       &   & - \frac{V}{N} \sum_{\bf k}
\frac{2 \cos(P_x - q_x - k_x) [\cos(P_y - q_y - k_y) - \cos(k_y)] F^{1/2}_4({\bf k}) }
{E - \varepsilon({\bf k}) - \varepsilon({\bf q}) - \varepsilon({\bf P}-{\bf q}-{\bf k})}    
\nonumber \\
       &   & - \frac{V}{N} \sum_{\bf k}
\frac{2 \cos(P_y - q_y - k_y) [\cos(P_y - q_y - k_y) - \cos(k_y)] F^{1/2}_5({\bf k}) }
{E - \varepsilon({\bf k}) - \varepsilon({\bf q}) - \varepsilon({\bf P}-{\bf q}-{\bf k})}    
\nonumber \\
       &   & - \frac{V}{N} \sum_{\bf k}
\frac{\cos(k_y) [\sin(k_x) - \sin(P_x - q_x - k_x)] F^{1/2}_6({\bf q}) }
{E - \varepsilon({\bf k}) - \varepsilon({\bf q}) - \varepsilon({\bf P}-{\bf q}-{\bf k})}    
\nonumber \\
       &   & - \frac{V}{N} \sum_{\bf k}
\frac{\cos(k_y) [\sin(k_y) - \sin(P_y - q_y - k_y)] F^{1/2}_7({\bf q}) }
{E - \varepsilon({\bf k}) - \varepsilon({\bf q}) - \varepsilon({\bf P}-{\bf q}-{\bf k})}    
\nonumber \\
       &   & - \frac{V}{N} \sum_{\bf k}
\frac{2 \sin(P_x - q_x - k_x) [\cos(P_y - q_y - k_y) - \cos(k_y)] F^{1/2}_8({\bf k}) }
{E - \varepsilon({\bf k}) - \varepsilon({\bf q}) - \varepsilon({\bf P}-{\bf q}-{\bf k})}    
\nonumber \\
       &   & - \frac{V}{N} \sum_{\bf k}
\frac{2 \sin(P_y - q_y - k_y) [\cos(P_y - q_y - k_y) - \cos(k_y)] F^{1/2}_9({\bf k}) }
{E - \varepsilon({\bf k}) - \varepsilon({\bf q}) - \varepsilon({\bf P}-{\bf q}-{\bf k})}  \: ,  
\label{3UV2d:eq:onethirtythree} 
\end{eqnarray}
\begin{eqnarray}
F^{1/2}_4({\bf q}) & = & \frac{U}{N} \sum_{\bf k} 
\frac{\cos(k_x) \, F^{1/2}_1({\bf q}) - \cos(k_x) \, F^{1/2}_1({\bf k})}
{E - \varepsilon({\bf k}) - \varepsilon({\bf q}) - \varepsilon({\bf P}-{\bf q}-{\bf k})}
\nonumber \\
       &   & - \frac{V}{N} \sum_{\bf k}
\frac{\cos(P_x - q_x - k_x) [\cos(P_x - q_x - k_x) - \cos(q_x)] F^{1/2}_2({\bf k}) }
{E - \varepsilon({\bf k}) - \varepsilon({\bf q}) - \varepsilon({\bf P}-{\bf q}-{\bf k})}    
\nonumber \\
       &   & - \frac{V}{N} \sum_{\bf k}
\frac{\cos(P_x - q_x - k_x) [\cos(P_y - q_y - k_y) - \cos(q_y)] F^{1/2}_3({\bf k}) }
{E - \varepsilon({\bf k}) - \varepsilon({\bf q}) - \varepsilon({\bf P}-{\bf q}-{\bf k})}    
\nonumber \\
       &   & - \frac{V}{N} \sum_{\bf k}
\frac{2 \cos^2(k_x) \, F^{1/2}_4({\bf q}) - 2 \cos(k_x) \cos(q_x) \, F^{1/2}_4({\bf k}) }
{E - \varepsilon({\bf k}) - \varepsilon({\bf q}) - \varepsilon({\bf P}-{\bf q}-{\bf k})}    
\nonumber \\
       &   & - \frac{V}{N} \sum_{\bf k}
\frac{2 \cos(k_x) \cos(k_y) \, F^{1/2}_5({\bf q}) - 2 \cos(k_x) \cos(q_y) \, F^{1/2}_5({\bf k}) }
{E - \varepsilon({\bf k}) - \varepsilon({\bf q}) - \varepsilon({\bf P}-{\bf q}-{\bf k})}    
\nonumber \\
       &   & - \frac{V}{N} \sum_{\bf k}
\frac{\cos(P_x - q_x - k_x) [\sin(P_x - q_x - k_x) - \sin(q_x)] F^{1/2}_6({\bf k}) }
{E - \varepsilon({\bf k}) - \varepsilon({\bf q}) - \varepsilon({\bf P}-{\bf q}-{\bf k})}    
\nonumber \\
       &   & - \frac{V}{N} \sum_{\bf k}
\frac{\cos(P_x - q_x - k_x) [\sin(P_y - q_y - k_y) - \sin(q_y)] F^{1/2}_7({\bf k}) }
{E - \varepsilon({\bf k}) - \varepsilon({\bf q}) - \varepsilon({\bf P}-{\bf q}-{\bf k})}    
\nonumber \\
       &   & - \frac{V}{N} \sum_{\bf k}
\frac{2 \cos(k_x) \sin(k_x) \, F^{1/2}_8({\bf q}) - 2 \cos(k_x) \sin(q_x) \, F^{1/2}_8({\bf k}) }
{E - \varepsilon({\bf k}) - \varepsilon({\bf q}) - \varepsilon({\bf P}-{\bf q}-{\bf k})}    
\nonumber \\
       &   & - \frac{V}{N} \sum_{\bf k}
\frac{2 \cos(k_x) \sin(k_y) \, F^{1/2}_9({\bf q}) - 2 \cos(k_x) \sin(q_y) \, F^{1/2}_9({\bf k}) }
{E - \varepsilon({\bf k}) - \varepsilon({\bf q}) - \varepsilon({\bf P}-{\bf q}-{\bf k})}  \: ,  
\label{3UV2d:eq:onethirtyfour} 
\end{eqnarray}
\begin{eqnarray}
F^{1/2}_5({\bf q}) & = & \frac{U}{N} \sum_{\bf k} 
\frac{\cos(k_y) \, F^{1/2}_1({\bf q}) - \cos(k_y) \, F^{1/2}_1({\bf k})}
{E - \varepsilon({\bf k}) - \varepsilon({\bf q}) - \varepsilon({\bf P}-{\bf q}-{\bf k})}
\nonumber \\
       &   & - \frac{V}{N} \sum_{\bf k}
\frac{\cos(P_y - q_y - k_y) [\cos(P_x - q_x - k_x) - \cos(q_x)] F^{1/2}_2({\bf k}) }
{E - \varepsilon({\bf k}) - \varepsilon({\bf q}) - \varepsilon({\bf P}-{\bf q}-{\bf k})}    
\nonumber \\
       &   & - \frac{V}{N} \sum_{\bf k}
\frac{\cos(P_y - q_y - k_y) [\cos(P_y - q_y - k_y) - \cos(q_y)] F^{1/2}_3({\bf k}) }
{E - \varepsilon({\bf k}) - \varepsilon({\bf q}) - \varepsilon({\bf P}-{\bf q}-{\bf k})}    
\nonumber \\
       &   & - \frac{V}{N} \sum_{\bf k}
\frac{2 \cos(k_y) \cos(k_x) \, F^{1/2}_4({\bf q}) - 2 \cos(k_y) \cos(q_x) \, F^{1/2}_4({\bf k}) }
{E - \varepsilon({\bf k}) - \varepsilon({\bf q}) - \varepsilon({\bf P}-{\bf q}-{\bf k})}    
\nonumber \\
       &   & - \frac{V}{N} \sum_{\bf k}
\frac{2 \cos^2(k_y) \, F^{1/2}_5({\bf q}) - 2 \cos(k_y) \cos(q_y) \, F^{1/2}_5({\bf k}) }
{E - \varepsilon({\bf k}) - \varepsilon({\bf q}) - \varepsilon({\bf P}-{\bf q}-{\bf k})}    
\nonumber \\
       &   & - \frac{V}{N} \sum_{\bf k}
\frac{\cos(P_y - q_y - k_y) [\sin(P_x - q_x - k_x) - \sin(q_x)] F^{1/2}_6({\bf k}) }
{E - \varepsilon({\bf k}) - \varepsilon({\bf q}) - \varepsilon({\bf P}-{\bf q}-{\bf k})}    
\nonumber \\
       &   & - \frac{V}{N} \sum_{\bf k}
\frac{\cos(P_y - q_y - k_y) [\sin(P_y - q_y - k_y) - \sin(q_y)] F^{1/2}_7({\bf k}) }
{E - \varepsilon({\bf k}) - \varepsilon({\bf q}) - \varepsilon({\bf P}-{\bf q}-{\bf k})}    
\nonumber \\
       &   & - \frac{V}{N} \sum_{\bf k}
\frac{2 \cos(k_y) \sin(k_x) \, F^{1/2}_8({\bf q}) - 2 \cos(k_y) \sin(q_x) \, F^{1/2}_8({\bf k}) }
{E - \varepsilon({\bf k}) - \varepsilon({\bf q}) - \varepsilon({\bf P}-{\bf q}-{\bf k})}    
\nonumber \\
       &   & - \frac{V}{N} \sum_{\bf k}
\frac{2 \cos(k_y) \sin(k_y) \, F^{1/2}_9({\bf q}) - 2 \cos(k_y) \sin(q_y) \, F^{1/2}_9({\bf k}) }
{E - \varepsilon({\bf k}) - \varepsilon({\bf q}) - \varepsilon({\bf P}-{\bf q}-{\bf k})}  \: ,  
\label{3UV2d:eq:onethirtyfive} 
\end{eqnarray}
\begin{eqnarray}
F^{1/2}_6({\bf q}) & = & \frac{U}{N} \sum_{\bf k} 
\frac{[\sin(P_x - q_x - k_x) - \sin(k_x)] F^{1/2}_1({\bf k})}
{E - \varepsilon({\bf k}) - \varepsilon({\bf q}) - \varepsilon({\bf P}-{\bf q}-{\bf k})}
\nonumber \\
       &   & - \frac{V}{N} \sum_{\bf k}
\frac{\sin(k_x) [\cos(k_x) - \cos(P_x - q_x - k_x)] F^{1/2}_2({\bf q}) }
{E - \varepsilon({\bf k}) - \varepsilon({\bf q}) - \varepsilon({\bf P}-{\bf q}-{\bf k})}    
\nonumber \\
       &   & - \frac{V}{N} \sum_{\bf k}
\frac{\sin(k_x) [\cos(k_y) - \cos(P_y - q_y - k_y)] F^{1/2}_3({\bf q}) }
{E - \varepsilon({\bf k}) - \varepsilon({\bf q}) - \varepsilon({\bf P}-{\bf q}-{\bf k})}    
\nonumber \\
       &   & - \frac{V}{N} \sum_{\bf k}
\frac{2 \cos(P_x - q_x - k_x) [\sin(P_x - q_x - k_x) - \sin(k_x)] F^{1/2}_4({\bf k}) }
{E - \varepsilon({\bf k}) - \varepsilon({\bf q}) - \varepsilon({\bf P}-{\bf q}-{\bf k})}    
\nonumber \\
       &   & - \frac{V}{N} \sum_{\bf k}
\frac{2 \cos(P_y - q_y - k_y) [\sin(P_x - q_x - k_x) - \sin(k_x)] F^{1/2}_5({\bf k}) }
{E - \varepsilon({\bf k}) - \varepsilon({\bf q}) - \varepsilon({\bf P}-{\bf q}-{\bf k})}    
\nonumber \\
       &   & - \frac{V}{N} \sum_{\bf k}
\frac{\sin(k_x) [\sin(k_x) - \sin(P_x - q_x - k_x)] F^{1/2}_6({\bf q}) }
{E - \varepsilon({\bf k}) - \varepsilon({\bf q}) - \varepsilon({\bf P}-{\bf q}-{\bf k})}    
\nonumber \\
       &   & - \frac{V}{N} \sum_{\bf k}
\frac{\sin(k_x) [\sin(k_y) - \sin(P_y - q_y - k_y)] F^{1/2}_7({\bf q}) }
{E - \varepsilon({\bf k}) - \varepsilon({\bf q}) - \varepsilon({\bf P}-{\bf q}-{\bf k})}    
\nonumber \\
       &   & - \frac{V}{N} \sum_{\bf k}
\frac{2 \sin(P_x - q_x - k_x) [\sin(P_x - q_x - k_x) - \sin(k_x)] F^{1/2}_8({\bf k}) }
{E - \varepsilon({\bf k}) - \varepsilon({\bf q}) - \varepsilon({\bf P}-{\bf q}-{\bf k})}    
\nonumber \\
       &   & - \frac{V}{N} \sum_{\bf k}
\frac{2 \sin(P_y - q_y - k_y) [\sin(P_x - q_x - k_x) - \sin(k_x)] F^{1/2}_9({\bf k}) }
{E - \varepsilon({\bf k}) - \varepsilon({\bf q}) - \varepsilon({\bf P}-{\bf q}-{\bf k})}  \: ,  
\label{3UV2d:eq:onethirtysix} 
\end{eqnarray}
\begin{eqnarray}
F^{1/2}_7({\bf q}) & = & \frac{U}{N} \sum_{\bf k} 
\frac{[\sin(P_y - q_y - k_y) - \sin(k_y)] F^{1/2}_1({\bf k})}
{E - \varepsilon({\bf k}) - \varepsilon({\bf q}) - \varepsilon({\bf P}-{\bf q}-{\bf k})}
\nonumber \\
       &   & - \frac{V}{N} \sum_{\bf k}
\frac{\sin(k_y) [\cos(k_x) - \cos(P_x - q_x - k_x)] F^{1/2}_2({\bf q}) }
{E - \varepsilon({\bf k}) - \varepsilon({\bf q}) - \varepsilon({\bf P}-{\bf q}-{\bf k})}    
\nonumber \\
       &   & - \frac{V}{N} \sum_{\bf k}
\frac{\sin(k_y) [\cos(k_y) - \cos(P_y - q_y - k_y)] F^{1/2}_3({\bf q}) }
{E - \varepsilon({\bf k}) - \varepsilon({\bf q}) - \varepsilon({\bf P}-{\bf q}-{\bf k})}    
\nonumber \\
       &   & - \frac{V}{N} \sum_{\bf k}
\frac{2 \cos(P_x - q_x - k_x) [\sin(P_y - q_y - k_y) - \sin(k_y)] F^{1/2}_4({\bf k}) }
{E - \varepsilon({\bf k}) - \varepsilon({\bf q}) - \varepsilon({\bf P}-{\bf q}-{\bf k})}    
\nonumber \\
       &   & - \frac{V}{N} \sum_{\bf k}
\frac{2 \cos(P_y - q_y - k_y) [\sin(P_y - q_y - k_y) - \sin(k_y)] F^{1/2}_5({\bf k}) }
{E - \varepsilon({\bf k}) - \varepsilon({\bf q}) - \varepsilon({\bf P}-{\bf q}-{\bf k})}    
\nonumber \\
       &   & - \frac{V}{N} \sum_{\bf k}
\frac{\sin(k_y) [\sin(k_x) - \sin(P_x - q_x - k_x)] F^{1/2}_6({\bf q}) }
{E - \varepsilon({\bf k}) - \varepsilon({\bf q}) - \varepsilon({\bf P}-{\bf q}-{\bf k})}    
\nonumber \\
       &   & - \frac{V}{N} \sum_{\bf k}
\frac{\sin(k_y) [\sin(k_y) - \sin(P_y - q_y - k_y)] F^{1/2}_7({\bf q}) }
{E - \varepsilon({\bf k}) - \varepsilon({\bf q}) - \varepsilon({\bf P}-{\bf q}-{\bf k})}    
\nonumber \\
       &   & - \frac{V}{N} \sum_{\bf k}
\frac{2 \sin(P_x - q_x - k_x) [\sin(P_y - q_y - k_y) - \sin(k_y)] F^{1/2}_8({\bf k}) }
{E - \varepsilon({\bf k}) - \varepsilon({\bf q}) - \varepsilon({\bf P}-{\bf q}-{\bf k})}    
\nonumber \\
       &   & - \frac{V}{N} \sum_{\bf k}
\frac{2 \sin(P_y - q_y - k_y) [\sin(P_y - q_y - k_y) - \sin(k_y)] F^{1/2}_9({\bf k}) }
{E - \varepsilon({\bf k}) - \varepsilon({\bf q}) - \varepsilon({\bf P}-{\bf q}-{\bf k})}  \: ,  
\label{3UV2d:eq:onethirtyseven} 
\end{eqnarray}
\begin{eqnarray}
F^{1/2}_8({\bf q}) & = & \frac{U}{N} \sum_{\bf k} 
\frac{\sin(k_x) \, F^{1/2}_1({\bf q}) - \sin(k_x) \, F^{1/2}_1({\bf k})}
{E - \varepsilon({\bf k}) - \varepsilon({\bf q}) - \varepsilon({\bf P}-{\bf q}-{\bf k})}
\nonumber \\
       &   & - \frac{V}{N} \sum_{\bf k}
\frac{\sin(P_x - q_x - k_x) [\cos(P_x - q_x - k_x) - \cos(q_x)] F^{1/2}_2({\bf k}) }
{E - \varepsilon({\bf k}) - \varepsilon({\bf q}) - \varepsilon({\bf P}-{\bf q}-{\bf k})}    
\nonumber \\
       &   & - \frac{V}{N} \sum_{\bf k}
\frac{\sin(P_x - q_x - k_x) [\cos(P_y - q_y - k_y) - \cos(q_y)] F^{1/2}_3({\bf k}) }
{E - \varepsilon({\bf k}) - \varepsilon({\bf q}) - \varepsilon({\bf P}-{\bf q}-{\bf k})}    
\nonumber \\
       &   & - \frac{V}{N} \sum_{\bf k}
\frac{2 \sin(k_x) \cos(k_x) \, F^{1/2}_4({\bf q}) - 2 \sin(k_x) \cos(q_x) \, F^{1/2}_4({\bf k}) }
{E - \varepsilon({\bf k}) - \varepsilon({\bf q}) - \varepsilon({\bf P}-{\bf q}-{\bf k})}    
\nonumber \\
       &   & - \frac{V}{N} \sum_{\bf k}
\frac{2 \sin(k_x) \cos(k_y) \, F^{1/2}_5({\bf q}) - 2 \sin(k_x) \cos(q_y) \, F^{1/2}_5({\bf k}) }
{E - \varepsilon({\bf k}) - \varepsilon({\bf q}) - \varepsilon({\bf P}-{\bf q}-{\bf k})}    
\nonumber \\
       &   & - \frac{V}{N} \sum_{\bf k}
\frac{\sin(P_x - q_x - k_x) [\sin(P_x - q_x - k_x) - \sin(q_x)] F^{1/2}_6({\bf k}) }
{E - \varepsilon({\bf k}) - \varepsilon({\bf q}) - \varepsilon({\bf P}-{\bf q}-{\bf k})}    
\nonumber \\
       &   & - \frac{V}{N} \sum_{\bf k}
\frac{\sin(P_x - q_x - k_x) [\sin(P_y - q_y - k_y) - \sin(q_y)] F^{1/2}_7({\bf k}) }
{E - \varepsilon({\bf k}) - \varepsilon({\bf q}) - \varepsilon({\bf P}-{\bf q}-{\bf k})}    
\nonumber \\
       &   & - \frac{V}{N} \sum_{\bf k}
\frac{2 \sin^2(k_x) \, F^{1/2}_8({\bf q}) - 2 \sin(k_x) \sin(q_x) \, F^{1/2}_8({\bf k}) }
{E - \varepsilon({\bf k}) - \varepsilon({\bf q}) - \varepsilon({\bf P}-{\bf q}-{\bf k})}    
\nonumber \\
       &   & - \frac{V}{N} \sum_{\bf k}
\frac{2 \sin(k_x) \sin(k_y) \, F^{1/2}_9({\bf q}) - 2 \sin(k_x) \sin(q_y) \, F^{1/2}_9({\bf k}) }
{E - \varepsilon({\bf k}) - \varepsilon({\bf q}) - \varepsilon({\bf P}-{\bf q}-{\bf k})}  \: ,  
\label{3UV2d:eq:onethirtyeight} 
\end{eqnarray}
\begin{eqnarray}
F^{1/2}_9({\bf q}) & = & \frac{U}{N} \sum_{\bf k} 
\frac{\sin(k_y) \, F^{1/2}_1({\bf q}) - \sin(k_y) \, F^{1/2}_1({\bf k})}
{E - \varepsilon({\bf k}) - \varepsilon({\bf q}) - \varepsilon({\bf P}-{\bf q}-{\bf k})}
\nonumber \\
       &   & - \frac{V}{N} \sum_{\bf k}
\frac{\sin(P_y - q_y - k_y) [\cos(P_x - q_x - k_x) - \cos(q_x)] F^{1/2}_2({\bf k}) }
{E - \varepsilon({\bf k}) - \varepsilon({\bf q}) - \varepsilon({\bf P}-{\bf q}-{\bf k})}    
\nonumber \\
       &   & - \frac{V}{N} \sum_{\bf k}
\frac{\sin(P_y - q_y - k_y) [\cos(P_y - q_y - k_y) - \cos(q_y)] F^{1/2}_3({\bf k}) }
{E - \varepsilon({\bf k}) - \varepsilon({\bf q}) - \varepsilon({\bf P}-{\bf q}-{\bf k})}    
\nonumber \\
       &   & - \frac{V}{N} \sum_{\bf k}
\frac{2 \sin(k_y) \cos(k_x) \, F^{1/2}_4({\bf q}) - 2 \sin(k_y) \cos(q_x) \, F^{1/2}_4({\bf k}) }
{E - \varepsilon({\bf k}) - \varepsilon({\bf q}) - \varepsilon({\bf P}-{\bf q}-{\bf k})}    
\nonumber \\
       &   & - \frac{V}{N} \sum_{\bf k}
\frac{2 \sin(k_y) \cos(k_y) \, F^{1/2}_5({\bf q}) - 2 \sin(k_y) \cos(q_y) \, F^{1/2}_5({\bf k}) }
{E - \varepsilon({\bf k}) - \varepsilon({\bf q}) - \varepsilon({\bf P}-{\bf q}-{\bf k})}    
\nonumber \\
       &   & - \frac{V}{N} \sum_{\bf k}
\frac{\sin(P_y - q_y - k_y) [\sin(P_x - q_x - k_x) - \sin(q_x)] F^{1/2}_6({\bf k}) }
{E - \varepsilon({\bf k}) - \varepsilon({\bf q}) - \varepsilon({\bf P}-{\bf q}-{\bf k})}    
\nonumber \\
       &   & - \frac{V}{N} \sum_{\bf k}
\frac{\sin(P_y - q_y - k_y) [\sin(P_y - q_y - k_y) - \sin(q_y)] F^{1/2}_7({\bf k}) }
{E - \varepsilon({\bf k}) - \varepsilon({\bf q}) - \varepsilon({\bf P}-{\bf q}-{\bf k})}    
\nonumber \\
       &   & - \frac{V}{N} \sum_{\bf k}
\frac{2 \sin(k_y) \sin(k_x) \, F^{1/2}_8({\bf q}) - 2 \sin(k_y) \sin(q_x) \, F^{1/2}_8({\bf k}) }
{E - \varepsilon({\bf k}) - \varepsilon({\bf q}) - \varepsilon({\bf P}-{\bf q}-{\bf k})}    
\nonumber \\
       &   & - \frac{V}{N} \sum_{\bf k}
\frac{2 \sin^2(k_y) \, F^{1/2}_9({\bf q}) - 2 \sin(k_y) \sin(q_y) \, F^{1/2}_9({\bf k}) }
{E - \varepsilon({\bf k}) - \varepsilon({\bf q}) - \varepsilon({\bf P}-{\bf q}-{\bf k})}  \: .  
\label{3UV2d:eq:onethirtynine} 
\end{eqnarray}

\vspace{1.0cm}

{\em Two-fermion problem. Singlet case.} The singlet spectrum $E({\bf K}) < 0$ is determined from the $3 \times 3$ secular determinant
\begin{equation}
\left\vert \begin{array}{ccc}
U M_{00} + 1  & - 2 V M_{10}                 & - 2 V M_{01}                \\
U M_{10}      & - V ( M_{20} + M_{00} ) + 1  & - 2 V M_{11}                \\
U M_{01}      & - 2 V M_{11}                 & - V ( M_{02} + M_{00} ) + 1  
\end{array} \right\vert = 0 \: ,
\label{3UV2d:eq:oneforty} 
\end{equation}
where 
\begin{equation}
M_{nm} \equiv \frac{1}{N} \sum_{\bf q} \frac{\cos{(nq_x)}\cos{(mq_y)}}
{\vert E \vert - a \cos{q_x} - b \cos{q_y}} \: ,  
\label{3UV2d:eq:onefortyone} 
\end{equation}
$a \equiv 4 t \cos{(K_x/2)} \geq 0$, $b \equiv 4 t \cos{(K_y/2)} \geq 0$, and $(K_x,K_y)$ are the components of the total lattice momentum of a pair ${\bf K}$. Although the integrals are well defined for $\vert E \vert > a + b$, and can always be computed numerically, searching for a minimal (pair plus free particle) energy at an arbitrary ${\bf P}$ requires many (thousands) evaluations of $M_{nm}$, which significantly slows down computation. It is much more efficient to use the following analytical expressions derived by standard but lengthy algebraic transformations:
\begin{equation}
M_{00} = \frac{2}{\pi \sqrt{\vert E \vert^2 - (a - b)^2 } } \: {\rm K}(k) \: ,  
\label{3UV2d:eq:onefortytwo} 
\end{equation}
\begin{equation}
M_{10} = \frac{2}{\pi a \sqrt{\vert E \vert^2 - (a - b)^2 } } 
\left\{ \left( \vert E \vert - b \right) {\rm K}(k) - \left( \vert E \vert - a - b \right)
\Pi\left( \frac{\pi}{2}; \frac{2a}{\vert E \vert + a - b}; k \right) 
\right\} \: ,  
\label{3UV2d:eq:onefortythree} 
\end{equation}
\begin{equation}
M_{01} = \frac{2}{\pi b \sqrt{\vert E \vert^2 - (a - b)^2 } } 
\left\{ \left( \vert E \vert - a \right) {\rm K}(k) - \left( \vert E \vert - a - b \right)
\Pi\left( \frac{\pi}{2}; \frac{2b}{\vert E \vert - a + b}; k \right) 
\right\} \: ,  
\label{3UV2d:eq:onefortyfour} 
\end{equation}
\begin{equation}
M_{11} = \frac{1}{\pi k \sqrt{ab} } \left\{ ( 2 - k^2) \: {\rm K}(k) - 2 {\rm E}(k) \right\} \: ,  
\label{3UV2d:eq:onefortyfive} 
\end{equation}
\begin{equation}
M_{20} = \frac{2}{\pi a^2 \sqrt{\vert E \vert^2 - (a - b)^2 } } 
\left\{ \left( \vert E \vert - b \right)^2 {\rm K}(k)    + 
\left[ \vert E \vert^2 - (a - b)^2 \right] \: {\rm E}(k) - 
2 \vert E \vert \left( \vert E \vert - a - b \right) 
\Pi\left( \frac{\pi}{2}; \frac{2a}{\vert E \vert + a - b}; k \right) 
\right\} ,  
\label{3UV2d:eq:onefortysix} 
\end{equation}
\begin{equation}
M_{02} = \frac{2}{\pi b^2 \sqrt{\vert E \vert^2 - (a - b)^2 } } 
\left\{ \left( \vert E \vert - a \right)^2 {\rm K}(k)    + 
\left[ \vert E \vert^2 - (a - b)^2 \right] \: {\rm E}(k) - 
2 \vert E \vert \left( \vert E \vert - a - b \right) 
\Pi\left( \frac{\pi}{2}; \frac{2b}{\vert E \vert - a + b}; k \right) 
\right\} ,  
\label{3UV2d:eq:onefortyseven} 
\end{equation}
\begin{equation}
k = \sqrt{\frac{4ab}{\vert E \vert^2 - (a - b)^2}} \: .  
\label{3UV2d:eq:onefortyeight} 
\end{equation}
Here ${\rm K}(k)$, ${\rm E}(k)$ and $\Pi(\cdots)$ are the complete elliptic integrals of the first, second and third types.

\vspace{1.0cm}

{\em Two-fermion problem. Triplet case.} The triplet spectrum factorizes into two $p$-states for {\em any} pair momentum ${\bf K}$ and can be determined from the two separate equations
\begin{equation}
V \left( M_{20} - M_{00} \right) + 1 = 0 \: ,  
\label{3UV2d:eq:onefortynine} 
\end{equation}
\begin{equation}
V \left( M_{02} - M_{00} \right) + 1 = 0 \: ,   
\label{3UV2d:eq:onefifty} 
\end{equation}
with the same expressions for $M_{nm}$ as in the singlet case. Along the Brillouin zone diagonal including ${\bf K} = (0,0)$, $a = b$, $M_{20} = M_{02}$, and the spectrum is doubly degenerate. In this case the integrals simplify to
\begin{equation}
M_{20} - M_{00} = M_{02} - M_{00} = \left\{ a = b \right\} = 
\frac{\vert E \vert}{a^2} \cdot \frac{2 {\rm E}(k) - \pi}{\pi} \: ,   
\label{3UV2d:eq:onefiftyone} 
\end{equation}
where $k = \frac{2a}{\vert E \vert}$.

\end{widetext}

%\bibliography{ThreeFermions_UV_2d} % Produces the bibliography via BibTeX.

%\end{comment}

\end{document}